\documentclass[twocolumn,showpacs,preprintnumbers,amsmath,amssymb]{revtex4}

\usepackage{graphicx}
\usepackage{dcolumn}
\usepackage{bm}
\usepackage{epsfig}

\begin{document}

\title{ Flexible Polyelectrolytes \\
with Monovalent Salt}
\author{Paulo S. Kuhn}
\affiliation{Departamento de F\'{\i}sica, Instituto de F\'{\i}sica e Matem\'atica\\ Universidade Federal de Pelotas, Pelotas, RS, Brasil}
\author{ Marcia C. Barbosa}
\affiliation{Instituto de F\'{\i}sica,\\ Universidade Federal do Rio Grande do Sul \\Caixa Postal 15051, 91501-970, Porto Alegre, RS, Brasil}
\email{barbosa@if.ufrgs.br}

\date{\today}
\begin{abstract}
We present a model for describing  flexible polyelectrolytes in a good solvent and in the presence of  monovalent salt .
The molecule composed by $N$ monomers is characterized by the end to end distance $R_e=b (Z-1)^\gamma$ and the number of associated counterions $n$. At high temperatures the polyelectrolyte behaves as a neutral polymer ($\gamma=0.588$). Decreasing the temperature, the macromolecule changes from this extended configuration($\gamma=0.588$)  to a stretched form ($\gamma\approx 1$). At even lower temperatures, above the Manning condensation threshold, the polyelectrolyte collapses ($\gamma\approx 0.3$). Our results show good agreement with 
simulations. 
\end{abstract}

\pacs{Valid PACS appear here}

\maketitle

\section{\label{sec1}Introduction}

Polyelectrolytes represent a very interesting class of materials. Biological
systems abound with polyelectrolytes. Two of the most well known of all 
polymers, DNA and RNA, are charged polymers  \cite{Ma78}\cite{Le02}. Besides, 
a large class of synthetic 
polyelectrolytes are present in the chemical industry. For instance, 
polyacrylic acid is the main ingredient for diapers \cite{Bu92} and 
dispersions of copolymers of acrylamide or methacrylamide and acrylic 
or methacrylic acid are fundamental for cleaning water  \cite{Sw77}.
Even though the tremendous interest in polyelectrolyte, unlike  neutral 
polymers \cite{Fl53}\cite{Ge79}, the understanding of the
behavior of electrically charged macromolecules is still rather poor.

The contrast between our understanding of charged and neutral polymers 
results from the long range nature of the electrostatic interactions that 
introduces new 
length  and time scales that render the 
analytical description very complicated \cite{Ba95}. The presence of charges 
introduces more 
than one new length scale making the scaling theories used for  no longer as 
simple and  applicable. Besides the length scale associated with the strength 
of the Coulomb interaction, there is 
also lengths associated with the mean separation of 
the counterions required by the charge neutrality. The association of 
these free ions to the polyion, renormalizing its charge,  is another factor 
that has to be 
taken into account.

Therefore, simulations seems to be a good technique to overcome these
major difficulties. Recent molecular simulations of salt-free 
systems \cite{St93}-\cite{Mi99} were able to obtain the end to end distance of
  strong electrolytes  below Manning parameter \cite{Manning2}. The 
picture provided by these simulations have shown to be more complicated 
than early analytic theories have predicted \cite{Od77}-\cite{Yethi}. 
Past theoretical works tended to neglect entropy.
For stiff chains, such as DNA, entropy is an small contribution
to the free energy and, in principle, can be disregarded. In contrast, for 
flexible polyelectrolytes, treating
entropy along with with the Coulomb interactions is essential. Acknowledging 
that, one of us
developed a theory for describing the thermodynamics of  salt-free 
polyelectrolytes in a good solvent. Within this approach  both entropy and 
electrostatic interactions 
are taken into account \cite{Ku02}\cite{Ku04}. The end to end distance 
and pressure calculated using this method show  
good agreement with simulations \cite{St93}\cite{St95}\cite{Wi98}. 

The addition of salt is a key factor both in biological and industrial 
applications. 
However, including  new charged species in the system, adds another
scale to this already very complicated problem. 
In this work we develop  a simple theory for studying a
dilute   polyelectrolyte solution  
in the presence of  monovalent salt in the framework of three
theories:
Debye and Bjerrum   for electrolytes \cite{DH1}-\cite{Bjerrum}, Manning 
condensation  \cite{Manning2} and Flory  elasticity for  polymers \cite{Fl53}.

Unlike neutral polymers, the polyelectrolyte in salt
solution exhibits three distinct behaviors.
At  high temperatures the chain is extended. As the temperature is decreased, the electrostatic repulsion between 
the monomers along the polyion becomes relevant and the chain  stretches.
At even lower temperatures,   above the Manning condensation threshold, 
counterions condense into the polyion. The  polymer chain net charge decreases
and the polyion collapses. The addition of salt screen the electrostatic 
interactions and the polymer contracts. 
Our results show a  good
quantitative agreement with simulations \cite{Stevens}.

 In section II,  our model is presented in
detail. In section III,  the free energy of  the system is constructed 
and minimized with respect to two parameters: number of ions associated
to the polyion and  end to end distance.  The results are presented  in 
section IV. Conclusions end this session.

\section{The model}

We consider a  dilute  polyelectrolyte solution of 
 concentration $\rho_p$ (see Figure 1). The chains are 
  immersed in a continuum solvent with dielectric constant 
$D$ and   monovalent salt.   
There are   $Z$ charged groups of diameter $\sigma$ along the 
polymer chain. 
$b=2^{1/6} \sigma$ is the distance between adjacent monomers. The 
total charge of the completely ionized molecule is  $-Zq<0$, where $q$ is 
the proton charge. The counterions that neutralizes the
solution have charge $+q$, diameter $\sigma$ 
and concentration $\rho_m = Z \rho_p$. 
The salt ions have charge $+q$ or $-q$,  diameter $\sigma$ and density $\rho_s$. For
simplicity the positive ions of salt will be refereed as counterions and 
the negative as coions. Thus
\begin{eqnarray}
\label{parameters}
\rho_+ = Z(1-m)\rho_p + \rho_s  \\
\rho_-=\rho_s
\end{eqnarray}
are the density of counterions and coions. The inverse Debye screening length 
$\kappa$ is defined in the usual way by 
$\kappa = \sqrt{4 \, \pi \, \rho_1 \, \lambda_B}$, where 
$\lambda_B = \beta \, q^2/D$ is the Bjerrum length, the distance 
between two ions where the electrostatic energy equals the thermal energy, 
$\rho_1 = \rho_+ + \rho_-$, and $\beta = 1/ k_B T$. 
The reduced density for species $j$ is $\rho^*_j = \rho_j \sigma^3$, the 
reduced temperature is $T^* = D \sigma/\beta q^2$. The Manning charge 
parameter is \cite{Manning2} 
$\xi = \beta \, q^2 / Db$. Hence $\xi = \sigma / b \, T^*$ and 
$\xi = \lambda_B/b$.
\begin{figure}
\begin{minipage}[b]{.8\linewidth}

\centering \epsfig{figure=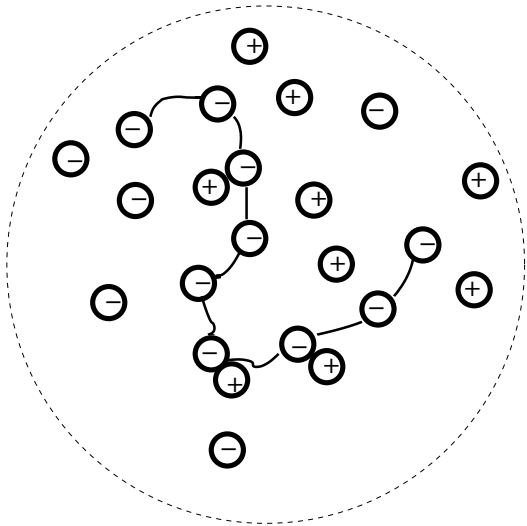,width=\linewidth}
\caption{The System}
\label{fig1}
\end{minipage} \hspace{.8cm}
\end{figure}
For simplicity,  macroion does not assume
  any preferential
geometry. Hence, the distance between monomers $i$ and $j$ 
is  represented by $r_{ij}=b \vert i-j \vert^{\gamma}$ as usual \cite{Fl53}. 
The end to end distance $R_e$ is the distance between monomers 
$1$ and $Z$ and it is represented by:
\begin{equation}
\label{Re}
R_e=b(Z-1)^{\gamma} \; .
\end{equation}

The electrostatic interaction between the chain and
the counterions leads the  formation of
a complexes made of 
one macromolecule and $n$ associated microions.
At a given temperature  $T$, monomer density $\rho_m=Z/V$ and salt 
density $\rho_s$, the system reaches the  equilibrium  characterized 
by a  number of associated counterions, $m$  and the end to end distance
characterized by $\gamma$.

\section{The Helmholtz free energy}

The system is, therefore, composed by complexes made 
of one polyion and $n$ counterions, free counterions and 
free coions. The end to end distance of the complex is 
$R_e$. The  equilibrium configuration is found by   minimizing  the
Helmholtz free energy density with respect to $n$ and $R_e$. The 
free energy density, $f=F/N_p$ is composed by three  contributions, namely
\begin{equation}  
\label{f}
f = f_{ELEC} + f_{HC} + f_{ENT} \,\,
\end{equation}
\noindent
where  $f_{ELEC}$ contains all  the electrostatic interactions, $f_{HC}$ 
has the hard core contribution between
the different species and $f_{ENT}$ has the entropic contributions for 
the free energy density.

The electrostatic free energy density is 
splited into:
\begin{equation}  
\label{fELEC}
f_{ELEC}=  f_{pc}+f_{ff} +f_{di}\,\,
\end{equation}
where   $f_{pc}$ accounts for
the electrostatic interaction between the 
polyelectrolyte chain and the free electrolyte, $f_{ff}$ takes care 
of the electrostatic interaction between the counterions e coions and $f_{di}$ includes the dipole-ion
interaction between the dipole pairs along the chain 
and  both charged monomers and free counterions and coions.

The hard core free energy density  is  divided into two contributions given by:
\begin{equation}  
\label{fHC}
f_{HC}=  f_{hc}+f_{cs} \,\,
\end{equation}
where  $f_{hc}$ takes care of  hard core interaction  between
 monomers along the
chain  and  $f_{cs}$ includes  
the hard core interaction between free ions. Here  $cs$ represents the
 free energy  of Carnahan-Starling, the hard core term between free ions.

Finally, the free energy density terms that contains entropic contributions 
 are:
\begin{equation}  
\label{fENT}
f_{ENT}= f_{el}+f_{ig}
\end{equation}
where  $f_{el}$ is the elastic  free
energy of the flexible chain, 
 $f_{ig}$ is the ideal gas free
energy for the mixture of the different species present in 
the solution. 

\subsection{The polyion-free ions electrostatic free energy density $f_{pc}$}

The electrostatic free energy density between the complex made 
of a polyelectrolyte and $n$ counterions associated to it and 
the free ions in the solution can be derived in the framework
of the Debye-Huckel theory yielding ( see appendix A):
\begin{equation} 
\label{fpc}
\beta f_{pc} = \frac {\xi \, p_z^2 \, b \, {\cal I}} {2} \, ,
\end{equation} 
where $p_z=-1+m=-1+Z/n$  and 
\begin{equation} 
\label{I}
{\cal I} = 2 \int^Z_0 dx \,(Z-x) \, \frac {e^{- \kappa r(x)}-1} {r(x)}
\end{equation} 
where $r(x)=b\;x^\gamma$.

These ideas have been successful in describing charged systems, including simple electrolyte solutions \cite{Levin1,Levin2}, charged rods 
\cite{YM1,Kuhn1},
charged spherical colloids \cite{YMM}, 
and flexible charged chains
\cite{Ku02,Ku04,Diehl}. In particular, the binding
isotherms of DNA with surfactant molecules have been found in good agreement
with experiment \cite{Kuhn3}.

The integral ${\cal I}$ might also be writen 
 using the incomplete gamma function \cite{GR3}, namely
\begin{equation} 
\label{gammainc}
\gamma_{inc} (\alpha ,x)= \int^x_0 dt \, e^{-t} t^{\alpha -1} \, , \,\,
Re \, \alpha > 0 \, .
\end{equation} \noindent
The suffix $inc$ is used to avoid confusion with the exponent $\gamma$.
In this case, we get
\begin{eqnarray}
\label{Iinc}
\frac {{\cal I}} {2} & = & \frac {Z \kappa} {(\kappa b)^{1/\gamma} \gamma}
\gamma_{inc} (-1+1/\gamma ,\kappa b Z^{\gamma})
\\ \nonumber &-& \frac {\kappa} {(\kappa b)^{2/\gamma} \gamma}
\gamma_{inc} (-1+2/\gamma ,\kappa b Z^{\gamma}) \nonumber \\
& & - \frac {Z^{2- \gamma}} {b (1 - \gamma)}
+ \frac {Z^{2- \gamma}} {b (2- \gamma)} \, .
\end{eqnarray} \noindent

\subsection{Free ion-free ion  electrostatic free energy density$f_{ff}$}

For the interaction between free ions we employ  the Debye-H\"uckel free energy
for electrolytes \cite{DH1,DH2},
\begin{equation}
\label{fff}
\beta f_{ff} = - \frac {1} {4 \pi \rho^*_p} \left[ \ln (1 + \kappa \sigma) - \kappa \sigma + (\kappa \sigma)^2/2 \right] \, .
\end{equation} \noindent

\subsection{The dipole-ion electrostatic free energy density $f_{di}$}

When a counterion associate to a negative monomer of the 
polyelectrolyte, it  forms dipole. The interaction between these dipoles 
formed by association and the monopoles consisting of free counterions and 
coions and non-associated monomers along the chain is given by
the usual dipole-ion interaction originally derived for 
the coulomb gas and given by \cite{Levin1,Levin2}:
\begin{equation} 
\label{fdi}
\beta f_{di}  = - \, \frac {1} {T^*} \, Z \, m \,
(\sigma /a_2)^3 \, x^2 \, \omega_2(x) \,\,\, ,
\end{equation}
\noindent
where $x= \bar{\kappa} a_2$, 
$\bar{\kappa} \sigma = \sqrt{\frac {4 \pi (\rho^*_1 + \bar{\rho}^*_-)} {T^*}}$, $\bar{\rho}^*_- = Z \, (1-m) \, \sigma^3/V_e$, 
$V_e = 4 \, \pi R_e^3/3$, $a_2 = 1.1619 \sigma$, and
\begin{equation} 
\label{omega}
\omega_2(x) = 3 [\ln (1+x+x^2/3) - x + x^2/6]/x^4 \,\,\, .
\end{equation}
\noindent
Note that we use here $\bar{\kappa}$ in order to take the interactions 
between the dipoles formed on the chain and the charged monomers and 
free ions. It is essential for the attractive forces that lead to 
collapse of the chain at low temperatures. The density $\bar{\rho}^*_-$ is 
the density of charged monomers not associated in the volume $V_e$ occupied
 by the chain.

\subsection{The excluded volume free energy densities $f_{hc}$ and $f_{cs}$}

The polyelectrolyte will be consider to be in a good solvent.
Therefore,  hard-core repulsion between monomers is approximated by a virial
coefficient from Flory-de Gennes theory \cite{Fl53,Ge79,deGennes2,deGennes3},
\begin{equation}  
\label{fhc}
\beta f_{hc}  = \frac {Z} {2} \, W_1 \, \bar{\rho} \,\, ,
\end{equation}
\noindent
where $W_1 = 4 \pi \sigma^3/3$ is the second virial coefficient for
hard spheres of diameter $\sigma$, and $\bar{\rho} = Z/V_e$ with $V_e=4\pi R_e^3/3$. 

The hard-core repulsion between free ions is
approximated by the Carnahan-Starling free energy \cite{CS},
\begin{equation}  
\label{fcs}
\beta f_{cs}  = V \, \rho_1 \, y \, \frac {4-3y} {(1-y)^2} \,\, ,
\end{equation}
\noindent
where $y=\pi \rho_1^*/6$ is the volume fraction occupied by the free positive and negative ions.

\subsection{The elastic free energy density $f_{el}$}

We are considering here flexible polyelectrolytes.
The entropic contribution of the elastic  free energy density
 is the same of a neutral polymer and it is given by
\cite{Fl53,Diehl,Flory2,Allen,Hill2}
\begin{equation} 
\label{fel}
\beta f_{el}  = \frac {3} {2} \, (\alpha^2 - 1) - 3 \ln \alpha \,\, ,
\end{equation}
\noindent
where $\alpha = R_e/R_0$, where $R_0=b \sqrt{Z-1}$ is the end to end distance                                               
of a polymer in a $\Theta$ solvent.

\subsection{The ideal gas free energy density $f_{ig}$}

Another important entropic contribution is due to the
mixing of the different species, complexes, counterions and 
coions. The free energy density associated with the mixing of ideal particles is given by  \cite{Levin1,Levin2}:
\begin{eqnarray}  
\label{fig}
\beta f_{ig} &=& \frac{1}{\rho_p}[\rho_+\ln \rho_+^*\Lambda^3 -\rho_++\rho_-\ln \rho_-^*\Lambda^3 \\ \nonumber &-&\rho_-]+ \ln \rho_p\Lambda^3 -1+\beta f_{ig}^c
\end{eqnarray}
\noindent
where $\Lambda= h/\sqrt{2 \pi m k_B T}$ is the 
thermal wavelength.  $f_{ig}^c$ 
includes  the internal degrees of freedom of  the complex made of a 
polyelectrolyte and the associated ions and it is given by:
\begin{eqnarray}  
\label{figc}
\beta f_{ig}^c &=& Z(1-m)[\ln \frac{Z(1-m)}{V_e}\Lambda^3 -1]
\\ \nonumber &+&Zm[\ln \frac{Zm}{V_e \zeta_2}\Lambda^6 -1]-Z[\ln \frac{Z}{V_e}\Lambda^3 -1]+\beta f_{ig}^{int} \; .
\end{eqnarray}

Since the  monomers along the chain are mobile, the first two
brackets in Eq.~(\ref{figc}) account for the entropy of mixing
of charged monomers and dipoles along the chain. The third bracket
 discounts the overcounting  since the entropy of a neutral
polymer was already accounted in
Eq.~(\ref{fel}). The last term is the internal partition 
function of the complex in its internal reference frame. 
The internal partition function of the dipoles, $\zeta_2$, 
is given by:
\begin{equation}  
\label{zeta2}
\zeta_2 = 4 \pi \int_{\sigma}^{R_{Bj}} r^2 \, e^{\sigma/rT^*} \, dr \equiv
K(T) \,\, ,
\end{equation}
\noindent
where $R_{Bj}$ is chosen according to Bjerrum to be $\sigma / 2 T^*$,
the location of the minimum of the integrand. Thus the association constant is
different from zero for $T^*<0.5$. Above this temperature there is no counterion
association and no dipole
formation on the chain. We have then
\cite{Bjerrum,Levin1,Levin2}
\begin{equation}  
\label{K}
K(T) = 4 \pi \sigma^3 Q_0 e^{1/T^*} T^* \,\, ,
\end{equation}
\noindent
with
\begin{eqnarray}  
\label{Q0}
Q_0 &=& \frac {1} {6 {T^*}^4} e^{-1/T^*} [Ei(1/T^*) - Ei(2) + e^2]
\\ \nonumber &-& \frac {1} {6 T^*} \left(\frac {1} {{T^*}^2} + \frac {1} {T^*} + 2 \right) \,\, ,
\end{eqnarray}
\noindent

The internal partition function of a $n$-complex contains: one electrostatic
term due to the interaction between the monomers and associated ions and one  term 
due to the different ways $n$ ions can associate into Z monomers in a polyions.
The additions of these two parcels gives:
\begin{eqnarray} 
\label{figint}
\beta f_{ig}^{int}  &=& \xi \, p_z^2 \, b \, \sum^{Z-1}_{i=1} \, (Z-i) \, \frac {1} {r(i)}\\ \nonumber &+&Z \, m \, \ln \, m + Z \, (1-m) \, \ln \, (1-m) \,\, ,
\end{eqnarray}
\noindent
with $r(i)=b \, i^{\gamma}$.

\section{Results and Conclusions}

The behavior of the polyelectrolyte under the variation of temperature 
is illustrated in Figure 2.   At 
high temperatures ( low Bjerrum length) the end to end distance
approaches  the one of a neutral polymer in a good solvent and
$R_e=b(Z-1)^{0.588}$. For $Z=32$ and 
$b=2^{1/6} \sigma$  the end to end distance is  
given by $R_e\approx 8.45 \sigma$. As the temperature decreases, the 
electrostatic interactions become dominant,  the polymer stretches 
and the end to end distance is given by
$\gamma\approx 1$. Above the Manning condensation threshold, the 
 counterions associate to the  monomers along the chain, decreasing
the electrostatic interactions. The elastic
energy becomes dominant and the polymer collapses. 
Our results are compared with simulations \cite{Stevens}. The actually salt
density
used for that simulation is not specified in the manuscript but is within 
the interval  $\rho_s^*=0.001$ and $\rho_s^*=0.008$.

Figures $3$ and $4$ show  that for a given $Z$, the radius of end to end distance, $R_e$ depends only on the value of the bare inverse of screening length, $\kappa_0 =\sqrt{4\pi(\rho_m+2\rho_s)\lambda_B}$, and  not on the density of polyelectrolytes. 
In Figure $3$   the renormalized  end to end distance given by
\begin{equation}
\label{R}
R^*=\frac{R_e}{b(Z/32)^{0.588})} 
\end{equation}
is ploted against $\ln \kappa_0 \sigma$ for $Z=32,64$ for varius densities  and for $\lambda_B=0.83 \sigma$ ($\xi=0.75$). The decrease of the end to end distance is, in this case, due to the screending 
of the salt present in the solution. Note that here there is no
condensation. 
The renormalized end to end distance, $R^*$, was constructed as 
follows. $R$ was divided by the end to end distance of 
an equivalent  neutral polymer in a good solvent
$R_0\approx b Z^{0.588}$.  In order to keep both $Z=32$ and $Z=64$ in the same
scale, $N$ was also divided by $Z=32$. Our results agree with simulations performed in the scale \cite{Stevens}.

In Figure $4$ we show $R^*$ for   $\lambda_B=3.2\sigma$($\xi=2.9$). In this case, since the system is above the Manning threshold, condensation is present and
the polyelectrolyte should form a more compact form than for $\lambda_B=0.83 \sigma$ ($\xi=0.75$).
Comparison with Figure $3$ shows that for $\ln \kappa_0\sigma> -0.5$ 
larger $\lambda_B$ has larger $R$. Part of this effect is 
due to plotting versus
$\ln \kappa_0\sigma$. For the same $\rho_m$ and $\rho_s$, the
end to end distance is surely smaller for $\lambda_B=0.83\sigma$ than
for $\lambda_B=3.2\sigma$($\xi=2.9$) as indicated by Figure $2$.

\begin{figure}
\begin{minipage}[b]{.8\linewidth}

\centering \epsfig{figure=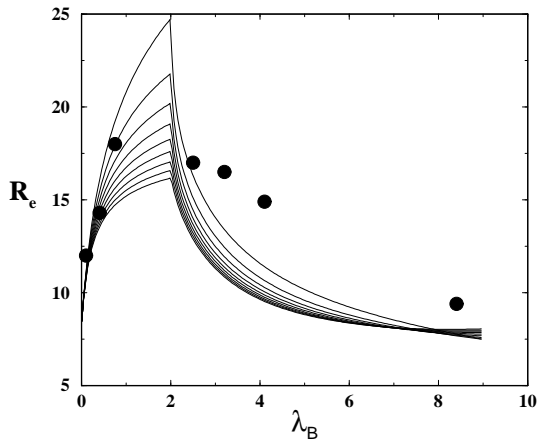,width=\linewidth}
\caption{Plot of end to end distance as a function of $\lambda_B$ for $Z=32$, 
 $\rho_m^*=0.001$,  and from top to bottom $\rho_s=0-0.008$. The circles  are the 
simulational results extracted from ref. \cite{Stevens}.}
\label{fig2}
\end{minipage} \hspace{.8cm}
\end{figure}

\begin{figure}
\begin{minipage}[b]{.8\linewidth}
\centering\epsfig{figure=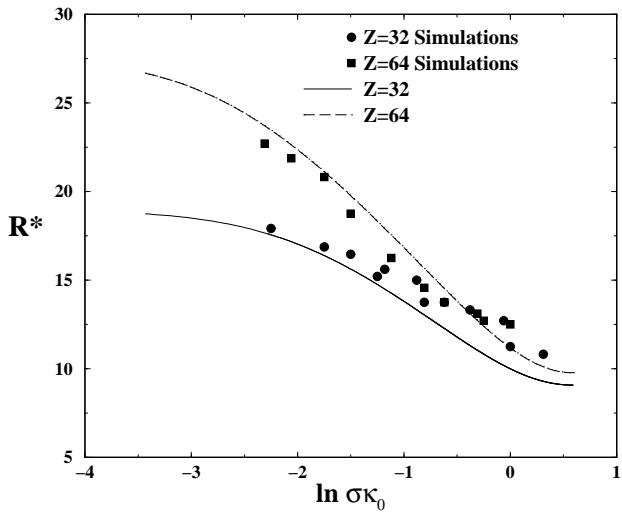,width=\linewidth,angle=-90}
\caption{Plot of the renormalized  end to end distance $R^*$ as a 
function of $\ln \kappa \sigma$ for $Z=32$ ( solid line ) and $Z=64$ ( dashed line), for  $\rho_m^*=0.0001,0.001,0.01,0.02$ and $\xi=0.75$. The points for different densities fall in the same line. The symbols are the simulation results for the same 
$\rho_m^*$ and $Z=32$ (circles) and $Z=64$ (squares) extracted from ref. \cite{Stevens}.}
\label{fig3}
\end{minipage} \hspace{.8cm}
\end{figure}

\begin{figure}
\begin{minipage}[b]{.8\linewidth}
\centering\epsfig{figure=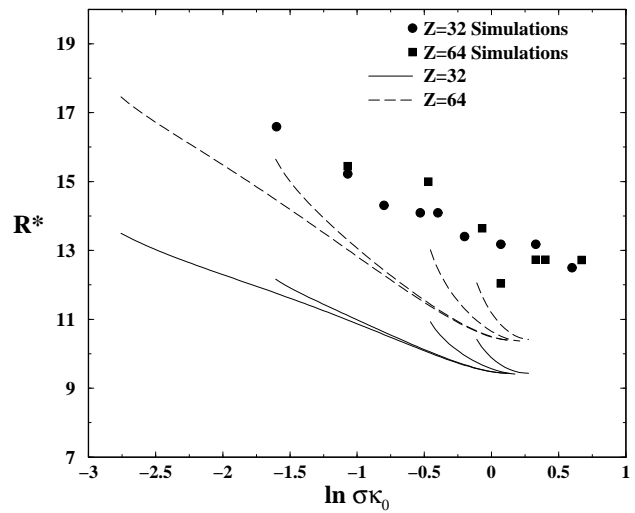,width=\linewidth,angle=-90}
\caption{Same as in Figure $3$ but for $\xi=2.9$.}
\label{fig4}
\end{minipage} \hspace{.8cm}
\end{figure}

In summary, we studied the end to end distance of  polyelectrolytes 
with added salt. We found that $R$ for a fixed $Z$ and $\lambda_B$ depends
only on the amount of salt in solution. In this case, the
 main effect of the electrolytes
is to screen the interactions and decreasing $R$. For fixed values of $Z$, 
$\rho_p$ and $\rho_s$, as the temperature is decreased, $R$ exhibits 
three distinct behaviors: stretched, extended and collapsed. This three regimes are related to the stretched polymer, fully extended polyelectrolyte and collapsed polymer behavior.
 
\begin{acknowledgments}

\vspace*{0.5cm} We thanks Paulo Krebs and Eduardo Henriques 
for the use of the computational 
resources at the Universidade Federal of Pelotas. This 
work was supported by the brazilian science agencies CNPq,
FINEP, and Fapergs.

\end{acknowledgments}

\appendix

\section{Calculation of $f_{pc}$}

In order to compute the free energy density associated
with the polyion-free ions interaction, let us start
by calculating the electrostatic potential at a point
$P$ due to the charged chains and electrolytes.
Each chain  has a linear charge density $\lambda ' = -(Z-n)q/L$.
The coordinates on the molecule are denote by a prime,
and the coordinates of the point $P$ where we want to 
evaluate the electrostatic potential are $P=(x,y,z)$.
The electrostatic potential at a point $P$ located in an
ionic solution, due to a charge element
$\lambda ' ds'$ on the molecule, is given by the Debye-H\"uckel
expression,
\begin{equation}  
\label{dphi}
d \phi= \frac { \lambda ' \, ds' \, e^{- \kappa r}} {Dr} \,\, ,
\end{equation}
\noindent
where $r$ is the distance of the charge element to the point $P$.
Integrating the expression above, the electrostatic total  
potential at $P$ becomes:
\begin{equation} 
\label{phi}
\phi (x,y,z) = \int d \phi = \frac {\lambda '} {D}
\int^{L}_0 ds' \frac {e^{- \kappa r}} {r} \,\, ,
\end{equation}
\noindent
where $L=b(Z-1)$ is the contour length. The electrostatic potential at 
$P$ due to the  chain when  there is no 
electrolyte  is given by:
\begin{equation} 
\label{phipr}
\phi_{pr} (x,y,z) = \int d \phi_{pr} = \frac {\lambda '} {D}
\int^{L}_0 ds' \frac {1} {r} \,\, ,
\end{equation}
\noindent
the suffix $pr$ meaning {\it proper}.

The potential  due to the ionic solution discounting the chain is given by
 $\psi \equiv \phi - \phi_{pr}$. The free energy is
 then evaluated by using  the charging process of Debye \cite{DH1,DH2} in
this potential.

The electrostatic energy of a charge element $dq$ at $(x,y,z)$ in the potential $\psi$ is
\begin{equation} 
\label{dU}
dU = dq(x,y,z) \, \psi (x,y,z) \,\, .
\end{equation} \noindent
The electrostatic energy of the entire molecule is obtained integrating 
the above expression, what gives
\begin{eqnarray} 
\label{U}
U & = & \frac {1} {2} \int dU =
\frac {1} {2} \int dq(x,y,z) \, \psi (x,y,z) = \frac {1} {2} \int \lambda \, \psi \, ds \, , \nonumber \\
& = & \frac {\lambda^2} {2D} \, \int^L_0 ds \, 
\int^{L}_0 ds' \frac {e^{- \kappa r}-1} {r} \, \,\, ,
\end{eqnarray}
 \noindent
where  $r=r(s-s')=r(s'-s)$ denotes the distance between line elements $ds$ and $ds'$, both on the molecule. The factor $1/2$ is included to avoid double counting. Now we substitute
$q$ by $\zeta q$ in $U$ and integrate in $\zeta$ from $0$ to $1$. We obtain
\begin{equation} 
F_{pc} = \int^1_0 \frac {2 U(\zeta) d \zeta} {\zeta}
 = 2 U \int^{1}_0 \zeta d \zeta = 2 U (\frac {1} {2} - 0)  = U \, .
\end{equation}
\noindent
Therefore,
\begin{equation} 
F_{pc} = U = \, \frac {\lambda^2} {2D} \, \int^L_0 ds \, 
\int^{L}_0 ds' \frac {e^{- \kappa r}-1} {r} \, \,\, .
\end{equation}
\noindent
We now change from $s$ to a dimensionless variable $t$, with
\begin{equation} 
L = \int^L_0 ds = \int^{Z}_0 b \, dt \, \,\, .
\end{equation}
\noindent
Hence,
\begin{equation} 
f_{pc} = \frac {\lambda^2 b^2} {2D} \, \int^{Z}_0 dt \, 
\int^{Z}_0 dt' \frac {e^{- \kappa r(t-t')}-1} {r(t-t')} \, \,\, .
\end{equation}
\noindent
Changing variables again, $x=t-t'$, and integrating by parts, the 
free energy becomes
\begin{equation} 
f_{pc} = \frac {\lambda^2 b^2} {2D} \, {\cal I} \, , \qquad
{\cal I} = 2 \int^Z_0 dx \,(Z-x) \, \frac {e^{- \kappa r(x)}-1} {r(x)} \, .
\end{equation}
 \noindent
Defining $p_z=-1+m$ and substituting $\lambda=(-1+m)q/b=p_z q/b$, we may write
\begin{equation} 
\beta f_{pc} = \frac {\xi \, p_z^2 \, b \, {\cal I}} {2} \, .
\end{equation} 


\noindent




\begin{thebibliography}{1}


\bibitem{Ma78} G. S. Manning, {\it Quart. Rev. Biophys.} {\bf II}(2),
179 (1978).


\bibitem{Le02} Y. Levin, {\it Rep. Prog. Phys.} {\bf 65}, 1577 (2002).

\bibitem{Bu92} F. Buchholz, {\it Trends Polym.}{\bf 99}, 277 (1992). 

\bibitem{Sw77}  SWIFT $\&$ CO, {\it  PURIFICATION OF WASTE WATER BY FLOTATION},
{\bf Patent number  GB1473481 } (1977)

\bibitem{Fl53} P. J. Flory, {\it Principles of Polymer Chemistry}
(Cornell University Press, Ithaca, New York, 1971).

\bibitem{Ge79} P. G. de Gennes, {\it Scaling Concepts in Polymer Physics}
(Cornell University Press, Ithaca, New York, 1979).

\bibitem{Ba95} J. -L. Barrat and J. -F. Joanny, {\it Adv. Chem. Phys.} {\bf 94}, 1 (1995).

\bibitem{St93} Mark J. Stevens and Kurt Kremer, {\it Phys. Rev. Lett.} {\bf 71}, 2228 (1993).

\bibitem{St95}  Mark J. Stevens and Kurt Kremer,{\it J. Chem. Phys. } {\bf 103}, 1669 (1995).

\bibitem{Mi99} U.  Micka, C. Holm and K. Kremer, {\it  Langmuir}{\bf 15}, 4033 (1999).

\bibitem{Manning2} G. S. Manning, {\it J. Chem. Phys.} {\bf 51}, 924 (1969).

\bibitem{Od77} T. Odijk, {\it J. Polym. Sci. Polym. Phys. Ed. } {\bf 15}, 477 (1977).

\bibitem{Sk77} J. Skolnick, M. Fixman, {\it Macromol.} {\bf  10}, 944 (1977).

\bibitem{Ba93} J. -L. Barrat and J. -F. Joanny, {\it Europhys. Lett.} {\bf 3}, 343 (1993).


\bibitem{Marcus} R. A. Marcus, {\it J. Chem. Phys.} {\bf 23}(6), 1057 (1955).



\bibitem{Katchalsky} A. Katchalsky, {\it Pure Appl. Chem.} {\bf 26}, 327 (1971).

\bibitem{Fixman} M. Fixman, {\it J. Chem. Phys.} {\bf 70}(11), 4995 (1979).

\bibitem{Odijk} T. Odijk, {\it J. Chem. Phys.} {\bf 93}(7), 5172 (1990).

\bibitem{Stigter} D. Stigter, {\it Biophys. J.} {\bf 69}, 380 (1995).

\bibitem{Bril} N. V. Brilliantov, D. V. Kuznetsov, R. Klein,
{\it Phys. Rev. Lett.} {\bf 81}(7), 1433 (1998).

\bibitem{Goles} R. Golestanian, M. Kardar, T. B. Liverpool
{\it Phys. Rev. Lett.} {\bf 82}(22), 4456 (1999).

\bibitem{Yethi} A. Yethiraj, {\it Phys. Rev. Lett.} {\bf 78}(19), 3789 (1997).


\bibitem{Ku02} P. S. Kuhn, {\it Physica A} {\bf 311}, 50 (2002).

\bibitem{Ku04}P. S. Kuhn, {\it Physica A} {\bf 337}, 481 (2004).


\bibitem{Wi98} R. G. Winkler, M. Gold, P. Reineker,
{\it Phys. Rev. Lett.} {\bf 80}(17), 3731 (1998).



\bibitem{DH1} P. W. Debye, E. H\"{u}ckel, {\it Phys. Z.} {\bf 24}, 185 (1923).

\bibitem{DH2} D. A. McQuarrie, {\it Statistical Mechanics}
(Harper and Row, New York, 1976), Cap. 15.

\bibitem{Bjerrum} N. Bjerrum, {\it Kgl. Dan. Vidensk. Selsk. Mat.-Fys. Medd.}
{\bf 7}, 1 (1926).


\bibitem{Stevens} M. J. Stevens, S. J. Plimpton,
{\it Eur. Phys. J.} {\bf B2}, 341 (1998).


\bibitem{Levin1} M. E. Fisher, Y. Levin, {\it Phys. Rev. Lett.}
{\bf 71} , 3826 (1993).

\bibitem{Levin2} Y.Levin, M. E. Fisher, {\it Physica A}
{\bf 225}, 164 (1996).

\bibitem{YM1}Y. Levin, M. C. Barbosa, {\it J. Phys. II (France)}
{\bf 7}, 37 (1997).



\bibitem{Kuhn1} P. S. Kuhn, Y. Levin, M. C. Barbosa,
{\it Macromolecules} {\bf 31} 5(4), 8347 (1998).

\bibitem{YMM} Y. Levin, M. C. Barbosa, M. N. Tamashiro,
{\it Europhys. Lett.} {\bf 41}, 123 (1998).

\bibitem{Diehl} A. Diehl, M. C. Barbosa, Y. Levin, {\it Phys. Rev.} {\bf E 54},
6516 (1996).







\bibitem{Kuhn3} P. S. Kuhn, Y. Levin, M. C. Barbosa,
{\it Chem. Phys. Lett.} {\bf 298}(4), 51 (1998).

\bibitem{GR3} I. S. Gradshteynm, I. M. Ryzhik, {\it Table of Integrals,
Series, and Products} (Academic Press, San Diego, 1994), p. 949.
\bibitem{deGennes2} P. G. de Gennes, {\it J. Phys. Lett.} {\bf 36}, L-55 (1975).

\bibitem{deGennes3} P. G. de Gennes, {\it Phys. Lett.} {\bf A 38}, 339 (1972).

\bibitem{CS} N. F. Carnahan, K. E. Starling, {\it J. Chem. Phys.} {\bf 51}(2),
635 (1969).

\bibitem{Flory2} P. J. Flory, {\it Proc. R. Soc. London} {\bf 351}, 351 (1976).



\bibitem{Allen} G. Allen, {\it Proc. R. Soc. London} {\bf 351}, 381 (1976).

\bibitem{Hill2} T. L. Hill, {\it An Introduction to Statistical Thermodynamics}
(Dover Publications, New York, 1986).



\end{thebibliography}
\end{document}